# Fifth giant ex-planet of the Solar System: characteristics and remnants

Yury I. Rogozin[*]

**Abstract**   In recent years it has coming to light that the early outer Solar System likely might have somewhat more planets than today. However, to date there is unknown what a former giant planet might in fact have represented and where its orbit may certainly have located. Using the originally suggested relations, we have found the reasonable orbital and physical characteristics of the icy giant ex-planet, which in the past may have orbited the Sun about in the halfway between Saturn and Uranus. Validity of the results obtained here is supported by a feasibility of these relations to other objects of the outer Solar System. A possible linkage between the fifth giant ex-planet and the puzzling objects of the outer Solar System such as the Saturn's rings and the irregular moons Triton and Phoebe existing today is briefly discussed.

**Keywords:** planets and satellites; individuals: fifth giant ex-planet, Saturn´s rings, Triton, Phoebe

**1 Introduction**

According to the existing ideas of the formation of the Solar System its planetary structure is held unchanged during about 4.5 billion years. Such a static situation of the things had been embodied, particularly, in offered in 1766 Titius-Bode's rule of the orbital distances for known at that time the seven planets from Mercury to Uranus. As it is known, the conformity to this rule for planets Neptune and Pluto discovered subsequently has appeared much worse than for before known seven planets. However, the essential departures of the real orbital distances from this rule for these two planets so far have not obtained any explained justifying within the framework of such conservative insights into a structure of the Solar System. But in this regard to the Solar System a certain progress has recently been achieved. Accordingly to the results of the recent computer simulation (Batygin et al 2012) the Solar System at the early stage of its dynamic evolution could have five outer planets (two gas giants and three icy giants). In other paper devoted to a possible fifth giant ex-planet (Nesvorný 2011) its author find that "the icy giants can form more easily at ‹ 15 AU from the Sun". Besides from, there by means of the computer simulation among other a case with its mass intermediate between that of Saturn and Uranus was tested. However, it is noteworthy, more appropriately the orbital and physical characteristics of this former fifth giant planet possibly at a rather recent epoch yet have not been determined and its possible remnants in the Solar System went unrecognized. Here, we state our approach to solve this problem and by that we try to give a key to unveiling some existing astronomical puzzles of the outer Solar System. In years of the planetary researches of the Solar System by the ground-based telescopes and *Voyager* and *Cassini* space missions a number of outstanding questions has been amassed on which as yet evidently there are no enough convincing answers. As it is known, the present-day planetary science has no an reasonably acceptable explanation for some puzzling astronomical facts, in particular, for that of an unambiguous origin of the rings and irregular moons of the giant planets, e. g., the largest retrograde moon of the Solar System Triton. In our opinion, to shed a new light on this situation the hypothesis for an existence in the past of an aforementioned outer giant planet is worthy of more detailed consideration here.

---

* Veda LLC, Moscow. Russian Federation, E-mail: yrogozin@gmail.com

## 2 Orbital characteristics of the fifth giant ex-planet

Possible existence in the past the fifth planet of the outer Solar System first was informed in our before published book (Yu. I. Rogozin "From numerical harmony to elementary astronomy", 2004, Moscow, Geos, ISBN 5-89118-359-5). As more adequate rule for the orbital distances of planets from the Sun than Titius-Bode's rule, which, as it is known, gives for Neptune and Pluto an error of 29 % and 95 %, respectively, compared to the observable quantities, we have found a new empirical rule for these distances. In so doing, a part of this rule relating to the outer planets being expressed by a dependence of the value a semi-major axis of an orbit of these planets $R_n$ on serial number of a planet $n$ is given by:

$$R_n/R_o = (n + 2)^2/ \Phi – 1/\Phi^2 \qquad (1),$$

where $R_n$, as usually, is expressed in astronomical units AU ($R_o$ = 1AU = 149.6 x $10^6$ km - distance from the Earth to the Sun), $\Phi$ = 1.618 – "factor of golden section", and $n$ = 1, 2 … respectively for Jupiter, Saturn etc. In case of replacement of ($n$ + 2) in Eq. (1) by $n$ with simultaneous change the former order of numbering by $n$ = 3, 4 … respectively for Jupiter, Saturn etc. this formula without considering the correction $1/\Phi^2$ shows up similar to the formula of electronic orbits in Bohr atomic model: $R_n \sim R_{no} n^2$, where $R_{no} = R_o /\Phi$. In so doing, the qualitative analogy of a structure, at least, of the outer Solar System and atomic structure is observed.

For the purposes of illustration of the similarity and distinctions to an available now picture of an arrangement of planets of the outer Solar System in the Table 1 the calculated by Eq. (1) and observed semi-major axes of these planets are given. To that it is possible to add that the orbital speed of the fifth giant ex-planet (informally nicknamed here Dispos (as reduced of "disrupted planet of the outer Solar System") could be equal to 7.674 km s$^{-1}$, and its orbital period could be equal to 58.5 years.

Table 1 Calculated and observed semi-major axes of five giant planets of the Solar System

| Planet name | Jupiter | Saturn | Dispos | Uranus | Neptune |
|---|---|---|---|---|---|
| Planet number, $n$ | 1 | 2 | 3 | 4 | 5 |
| $R_n$ (calculated), AU | 5.18 | 9.51 | 15.07 | 21.87 | 29.90 |
| $R_n$ (observed), AU | 5.20 | 9.54 | - | 19.19 | 30.07 |

Data of this table testify to an expected improvement in a fit between calculated and observed values of the semi-major axes of Neptune compared to Titius-Bode's rule. Also, it is appeared that compared to previously known rules of such a type our rule gives two totally new results that are as follows. First, in the past between planets Saturn and Uranus one more outer planet appropriate to number 3 in the Table 1 above-stated the planet Dispos, could has been in existence. Second, during an existence of the planet Dispos the orbit of Uranus should be much further from the Sun than at present epoch. On our sight, it would be logical to believe that the reason to occur the change in the orbital distance of Uranus of the Sun, accompanied by the strong tilt of its axis, could be a cardinal realignment of a gravitational interaction of the nearby planets of the outer Solar System, as a result of a disappearance the planet Dispos, if its mass was comparable with their masses. The reality of an existence in the past such a planet Dispos as a result of the disruption of what Uranus by the gravitational fields of Saturn and Jupiter has been pulled at inward the Sun apparently can be confirmed only by virtue of the determination of its physical characteristics and subsequent comparison with the appropriate ones of other planets of the outer Solar System.

## 2 Determination of the physical characteristics of giant ex-planet Dispos

In the context of present-day astrophysics the determination of the physical characteristics of non-existent in the present a planet appears to be an intractable problem. Nevertheless, we could yet find the appropriate method. Its first step was the determination of the density of the planet Dispos. In one of our former papers (The journal "The natural and engineering sciences" # 1, pp. 49-51, 2006, ISSN 1684-2626, Moscow) based on an analysis of the orbital data and the physical characteristics of known planets of the Solar System we have revealed a fact of the existance of the orbital parameters - density behavior for these planets. It is displayed by the semi-major axis $R$- dependence of the product of the density of planet $\rho$ by square of its orbital speed $v^2$, which has dimensions of the pressure (informally is named for the orbital pressure). Using known data for planets of the outer Solar System, a behavior of this dependence is available in the lower line of the Table 2.

Table 2 Orbital pressure data of the present outer planets of the Solar System

| Parameters | Jupiter | Saturn | Uranus | Neptune |
|---|---|---|---|---|
| Semi-major axis, $R_n$ (AU) | 5.2034 | 9.537 | 19.19 | 30.0689 |
| Mean orbital speed, $v$ (km s$^{-1}$) | 13.06 | 9.646 | 6.80 | 5.4334 |
| Mean density, $\rho$ (g cm$^{-3}$) | 1.326 | 0.687 | 1.27 | 1.638 |
| Orbital pressure, $\rho v^2$ (g cm$^{-3}$) (km s$^{-1}$) | 226 | 63.92 | 58.72 | 48.36 |

Note: used here unit of the orbital pressure 1(g cm$^{-3}$) (km s$^{-1}$)$^2$ = $10^9$ Pa

Based on a decreasing behavior of this function $F(R) = \rho v^2$ with an increase in $R$ the most suitable type of approximation of this function was chosen by:

$$F(R) = \rho v^2 = a R^b e^{Rc}, \qquad (2)$$

where $a > 0$ and at least one of factors $b$ or $c < 0$. Using tabulated data for the planets Saturn, Uranus (on old orbit) and Neptune ($R$ equals 9.537 AU, 21.87 AU, and 30.0689 AU, respectively; $v$ equals 9.646 km s$^{-1}$, 6.37 km s$^{-1}$, and 5.4334 km s$^{-1}$, respectively; and $\rho$ equals 0.687 g cm$^{-3}$, 1.27 g cm$^{-3}$, and 1.638 g cm$^{-3}$, respectively), when solving this system of three equations the following values of coefficients were obtained: $a = 119$, $b = -0.28562$, and $c = -0.0023786$. Substituting these values into Eq. (2) with the stated above value of the semi-major axis of this planet $R = 15.0689$ AU and the appropriate orbital speed $v = 7.674$ km s$^{-1}$ gives the density of the planet Dispos, which is equal to 0.963 g cm$^{-3}$ (that is close enough to the density of water ice of 0.92 g cm$^{-3}$). In absence of any evidences for the sizes of this planet the following step should be the determination of its mass. Based on the assumption about a primary harmony of Nature in general and the Solar System in particular, we shall attempt to reveal an existence of the functional dependence of the mass of the outer planets on their distance from the Sun. It has appeared, that similarly to the law of orbital pressure by Eq. (2) for the outer planets Jupiter, Saturn, and Uranus (at its former orbit) exists the law of orbital dependence of the angular momentum (as it show a respective check Neptune does not obey this dependence), i.e.:

$$M v R = a R^b e^{Rc}. \qquad (3)$$

Using the known parameters of three above-stated planets Jupiter, Saturn, and Uranus (mass $M$ equals 317.83, 95.16, and 14.5357 mass of the Earth, respectively; $v$ equals 13.06 km s$^{-1}$, 9.646 km s$^{-1}$, and 6.37 km s$^{-1}$, respectively, and $R$ equals 5.2034 AU, 9.537 AU, and 21.87 AU, respectively) for the function satisfying to Eq. (3), such values of its coefficients may have obtained: $a = 1.908 \times 10^5$, $b = -1.22484$, and $c = -0.03665$. The substitution the orbital data of the planet Dispos in the obtained equation gives its mass of 35.3525 mass of the Earth (or of $211.14 \times 10^{24}$ kg in terms of Earth's mass of $5.97237 \times 10^{24}$ kg). From the obtained values of its density and mass follows radius of this planet appears to be equal to 37407 km. As a result, the surface gravity of this planet may have been equal to 10.07 m s$^{-2}$. To clearly represent the found physical characteristics of the Dispos compared to the tabulated data of two near giant planets in the Solar System they are given in the Table 3.

Table 3 Comparative physical characteristics of planet Dispos and two nearby giant planets

| Planet name | Mean radius (km) | Mass ($10^{24}$ kg) | Density (g cm$^{-3}$) | Surface gravity (m s$^{-2}$) |
|---|---|---|---|---|
| Saturn | 58232 | 568.34 | 0.687 | 10.44 |
| Dispos | 37407 | 211.14 | 0.963 | 10.07 |
| Uranus | 25362 | 86.810 | 1.27 | 8.69 |

## 3 Remnants of the disrupted ex-planet Dispos in the outer Solar System

The disrupted fifth giant planet could not have disappeared apparently having not left behind any visible traces in space. As them there could have become, in particular, such strange objects of the outer Solar System as rings of the giant planets and retrograde moons of Neptune and Saturn. Below we must try to justify this assumption, having connected characteristics of these puzzling objects to that of the fifth giant ex-planet of the outer Solar System.

### 3. 1 On the origin of the rings of the outer giant planets

The orbital and physical characteristics of this disrupted planet found here apparently can be directly concerned with an origin of rings Saturn, Uranus, and Neptune. In our opinion it is necessary to explore an origin of Saturn´s rings in conjunction with an origin of the rings of Uranus and Neptune. Actually, the available scenario of the formation of Saturn's rings from the fragments of some disrupted Saturn's moon sized about 100 km can´t be extended to an origin of rings of two other giant planets, which are strongly distanced from so much lesser object. A collision of some assumed Saturn's moon with one of the asteroids in the present structure of the Solar System shows up an unbelievable event in view of the large remoteness of both asteroid belts from Saturn. As the history of the Comet of Shoemaker-Levy fall on Jupiter has been shown, the collision of one of the comets on any moon of the giant planets even of such large sizes as Galilean moons appear can´t too be a real possible reason for a disruption of so much lesser moon to form these rings. An explanation for a disappearance of a rock core of such a moon after a possible loss of its icy envelope to fall on Saturn also is unimpressive, as with one exception retrograde moon Triton other moons of the giant planets, as it is known, do not show the consecutive trend to bring into proximity with their hosts.

According to the value of its mean density it is possible to consider planet Dispos as basically icy planet. As Saturn's rings and those of other planets consist mainly of the particles of water ice and accordingly to the data of *Voyager* space mission for Saturn have young age, they could not have been formed simultaneously with a planet. Thus, it is possible to assume, that they have been appeared as a result of a disruption of fifth giant ex-planet and a subsequent drift of its remnants in the direction of the orbits other outer planets, where they could be captured by these planets. It is known, that by the results of the *Voyager* space mission the Saturn's rings could have a mere 100 million years old. One of the results of more recent *Cassini* mission (Hedman et al 2007) is that, as it was established, during the elapsed 25 years the image of the center of the D72 ringlet in the D ring has been shifted inwards by over 200 km. It seems implies the speed of a moving of the rings in direction to Saturn makes up of about 8 km yr$^{-1}$. In such a case the mean distance between these planets 15.069 AU - 9.537 AU = 5.532 AU = 827.6 x 10$^6$ km some remnants of the disrupted ex-planet could be overcome in a time of about 103.45 Myr. Thus, in view of the significant sizes of this planet and a possible wide scatter of its remnants in the space the process to thrown them into the Saturn's orbit, probably, could have last much thousand years, that probably is a possible reason for the longest existence of these rings and the basis to form new previously invisible faint Saturn's rings. Additionally, the more recent study (Zhang et al 2017) argues the young age as well of Saturn´s rings A and B less than 150 Myr although proposes also its origin from recent breakup of an icy moon a questionable probability of which we have just discussed above. In more recent paper presented on the results of the Cassini´s mission, which ended on September 15, 2017 (Cuzzi 2018) there was confirmed a young age Saturn´s rings on the order of 100 Myr ago, but a plausible explanation for the origin of the such young age of the rings, as was noted by the author, yet is lacking.

Below as the proposed key to this puzzle, we give the following rough assessment of the Saturn´s rings location as the possible remnants of the disrupted planet Dispos. The initial speed of the icy particles of the remnants of this planet at its surface was 27.081 km s$^{-1}$ as the sum of the orbital speed of planet of 7.674 km s$^{-1}$ and the speed of its surface particles motion around the center of the planet of 19.407 km s$^{-1}$. Ideally, when maintaining this orbital speed relative to the Sun, these icy particles being captured by the gravity of Saturn would have to sit in an orbit apart from its center by 51730 km whereas the present location of the nearest to Saturn the D ring is at a distance of 66900÷74510 km from its center. But due to the natural dynamic braking of the icy particles when their transit to Saturn at the expense of the multiple mutual collisions and an action of a number of other factors they could be reducing their orbital speed and are located at a greater distance from Saturn in their present positions.

## 3. 2 On the origin of the retrograde moons Triton and Phoebe

Other mysterious object of the outer Solar System is the retrograde moon of Neptune Triton. It is generally agreed it is a captured celestial body. As Triton is retrograde moon it could have been captured only on opposing motion with reference to the Neptune's orbiting the Sun. Based on almost all celestial bodies in the Solar System are orbiting in the same direction, for to have a retrograde motion Triton must have been sometime to gain a head-on impulse. In view of above reasoning on negligible probability of a collision of such celestial bodies as moons of the planets on other large celestial bodies, we have assumed that Triton could be gain the head-on impulse only in structure of the large disrupted icy planet Dispos. Considering its true spherical shape the only opportunity for Triton to preserve an intact surface at collision with any other celestial body could be its location only inside this planet in a moment of such a collision. The quite spherical shape of Triton suggests it could have been a core of the planet Dispos. One of the reasons for the benefit of such an assumption is a negligible surface age of Triton with its upper limit of ~ 50 Myr (Schenk et al 2007). Other basis for such an assumption is the relation between the Triton's the present density of 2.061g cm$^{-3}$ and the above mean density of the planet Dispos of 0.963 g cm$^{-3}$ that equals 2.140. This value is close to 2.139 that, as we believe, is general for the relation between the core's density and the mean density of planets of the Solar System (at least, it is quite obeyed for terrestrial planets the Earth, Mars, and Venus). Alongside with the widely known mathematical constants $\pi$ (3.14159), $\Phi$ (1.61803) and $e$ (2.71828) the constant, symbolized here as $\theta = 2.139$, is also the mathematical constant, which is conformed to the identity $\pi \Phi \theta = 4e$. In this connection it is possible to specify that based on data of the density of the inner and outer cores of the Earth (Anderson 1989) in whole the density of the core of the Earth $\rho_c$ is possible to estimate as approximately equal to 11.91 g cm$^{-3}$. The ratio of this value $\rho_c$ to the mean density of the Earth $\rho_o = 5.514$ g cm$^{-3}$ makes up 2.160 that differs from $\theta$ by only ~1 %. As to Venus, condition $\rho_c = 2.139 \rho_o$ is carried out at its core mass and radius of 20 % of total mass of the planet and of 2755 km, respectively. These numbers correlate well with the present estimates of these parameters of Venus' core. The described above possible origin of Triton as a core of the ex-planet Dispos appear could be also explain the extremely low average temperature of its surface of 38 K whereas Neptune has the surface temperature of 72 K.

We do not can know with certainty the true reason for a disruption of the planet Dispos. It is possible only to assume that as such a reason may have become its collision with one of the large comets. Of it an existence of strange retrograde Saturn's moon Phoebe can testify. Its unusual fragmental shape, which can be seen on snapshots obtained by *Cassini* space mission, is not similar to that of other moons of Saturn, which are in hydrostatic equilibrium, and more in the nature is a fragment knocked out from some heavenly body. In favor of this assumption the traces of a cometary origin found out at its surface and large quantities of ice below a relatively thin blanket of the dark surface deposits testify. A speed of this comet could be likely comprised 12.01 km s$^{-1}$ as a sum of the orbital speed of Dispos (7.674 km s$^{-1}$) and the initial speed of Triton orbited Neptune (4.389 km s$^{-1}$). With consideration for a possible compression as the result of a cometary impact, a mean density of Phoebe of 1.634 g cm$^{-3}$ could be reasonably expected to be intermediate one between the density of a core of the planet Dispos of 2.060 g cm$^{-3}$ and its mean density of 0.963 g cm$^{-3}$.

**Conclusions**

In this work we have tried to support a true existence of the fifth giant ex-planet not only in the early outer Solar System, as it has been shown recently (Nesvorný 2011, Batigin et al 2012), but in a rather recent epoch of about 100 Myr ago. Its calculated characteristics are in the harmonic line with those of the existing planetary objects of the outer Solar System. Based on the past existence proved here of such an icy giant planet we have offered also the new hypothesis for origins of the some puzzling objects of the outer Solar System as its retained remains. Besides, the validity of the orbital and physical characteristics of this giant ex-planet calculated here is confirmed by an existence of its direct relations to those of the present-day trans-Neptunian dwarf planets, in particular, Pluto, which we propose to present elsewhere.